\documentclass[preprint]{revtex4}
\usepackage{graphicx}
\usepackage{bm}
\usepackage{subfigure}

\begin{document}

\title{Radiative diagnostics for sub-Larmor scale magnetic turbulence}
\date{\today}

\author{Sarah J. \surname{Reynolds}}
\affiliation{Department of Physics and Astronomy, University of Kansas, Lawrence, KS 66045}
\author{Mikhail V. \surname{Medvedev}}
\affiliation{Department of Physics and Astronomy, University of Kansas, Lawrence, KS 66045}
\affiliation{Institute for Nuclear Fusion, RRC ``Kurchatov Institute'', Moscow 123182, Russia}

\begin{abstract}
Radiative diagnostics of high-energy density plasmas is addressed in this paper. We propose that the radiation produced by energetic particles in small-scale magnetic field turbulence, which can occur in laser-plasma experiments, collisionless shocks, and during magnetic reconnection, can be used to deduce some properties of the turbulent magnetic field.  Particles propagating through such turbulence encounter locally strong magnetic fields, but over lengths much shorter than a particle gyroradius (that is, $\lambda_B \ll \gamma mv/q\left\langle B_{\bot} \right\rangle$).  Consequently, the particle is accelerated but not deviated substantially from a straight line path.  We develop the general jitter radiation solutions for this case and show that the resulting radiation is \emph{directly} dependent upon the spectral distribution of the magnetic field through which the particle propagates.  We demonstrate the power of this approach in considering the radiation produced by particles moving through a region in which a (Weibel-like) filamentation instability grows magnetic fields randomly oriented in a plane transverse to counterstreaming particle populations.  We calculate the spectrum as would be seen from the original particle population and as could be seen by using a quasi-monoenergetic electron beam to probe the turbulent region at various angles to the filamentation axis.   

\end{abstract}

\pacs{}
\keywords{}

\maketitle

\section{\label{sec:intro}Introduction}

Understanding the dynamics of small-scale magnetic field turbulence is of great interest in both laboratory and astrophysical plasmas due to its strong effect on particle distributions, as well as its involvement in the spontaneous generation, evolution, and dissipation of magnetic field.  However, it remains difficult to relate the intense radiation produced by these sources to the underlying physical processes or field structure within them.  Calculating the spectrum of such radiation poses unique analytical challenges due to the magnetic field varying rapidly and down to very small scales.  Large-scale magnetic field structure and some details of the particle distribution in energy can be obtained through comparing spectra with synchrotron radiation; however, this fails where particle coupling to field turbulence is most intense.  Furthermore, while the instabilities that generate such turbulence are often seen laser-plasma experiments, they can easily be disrupted or altered by the presence of in situ probes, and thus the formation and evolution of the turbulent magnetic field remains difficult to study quantitatively.

The theory of jitter radiation, named for the particle's ``jitter'' as it undergoes a series of small transverse accelerations without a significant path deviation, has been developed \citep{mdv00, mdv06} to improve upon this by considering the radiation emitted by a relativistic charged particle moving through a magnetic field that varies on spatial scales much smaller than a typical particle's gyroradius. In this case the magnetic field correlation length $\lambda_B$ is too small for a particle of Lorentz factor $\gamma$ to be deflected beyond the opening angle $1/\gamma$ into which most of its radiation will be relativistically beamed.  This differs significantly from the synchrotron case in which the radiation spectrum is determined primarily by the sweep of the relativistic beaming cone past an observer's direction.  For a magnetic field with average strength $\left\langle B_{\perp}\right\rangle$ perpendicular to the direction of particle motion and correlation length $\lambda_B$, we can thus define a parameter $\delta_{rad}$ which determines the degree to which the radiation will be dominated by jitter or synchrotron radiation:

\begin{equation}\label{eq:jitterparameter}
\delta_{rad} = \gamma\alpha = \frac{e\left\langle B_{\perp} \right\rangle\lambda_B}{mc^2}
\end{equation}

For $\delta_{rad} \gg \gamma$, we have the synchrotron radiation case, in which the resulting spectrum is dominated by the frequency of the particles' redirection due to the large-scale magnetic field.  For $\delta_{rad} \ll 1$, we have small-angle jitter radiation, in which the emission into a particular direction is produced by particles which undergo accelerations but not significant deflections, and thus radiate into the same direction over many field correlation lengths.  In the intermediate region $1 \alt \delta_{rad} \alt \gamma$, the radiation spectrum is influenced both by the path geometry and the magnetic field distribution.  In such a case, which we designate as large-angle jitter radiation, the particle is deflected beyond the relativistic beaming angle, but the field geometry along the path segment before this deflection occurs leads to the occurence of low-frequency harmonics that differ from the synchrotron.  (See \cite{mdvdet} for more on both the small- and large-angle jitter regimes and associated spectral features.)   Undulator and ``wiggler'' radiation can be in the jitter regime, but are results for specific magnetic field configurations selected to coherently amplify certain harmonics, whereas the radiation produced by some distribution of particles moving in a region of turbulent plasma is generally an incoherent average over magnetic fields which randomly vary in both strength and orientation.  Distinguishing jitter radiation from synchrotron radiation by spectral features alone is by no means trivial.  However, for certain field configurations the jitter radiation spectrum can be notably harder at lower energies (below the spectral peak or break) than is possible from synchrotron radiation, which regardless of the particle distribution is limited to a low-energy spectral index of no greater than 1/3 (or -2/3 for the photon spectrum).  

While synchrotron radiation reflects primarily the statistical properties of the particle distribution, the jitter radiation spectrum will directly reflect the statistical properties of the magnetic field along a particular line-of-sight.  Jitter radiation is especially useful for experimental diagnostics because of its sensitivity to the magnetic field along a particular path and with a particular orientation to any field anisotropy.  An ideal experimental design would make use of this by using a quasimonoenergetic electron beam to probe the site of plasma turbulence at a variety of incident angles and measure the resulting radiation, which should be emitted primarily along the direction of the probe beam. The observed radiation spectrum can then be compared with the theoretically calculated jitter spectrum, through the equations we develop in section \ref{sec:theory}.  In section \ref{sec:results}, we apply these equations to calculate the spectrum for a sample application in which a quasimonoenergetic beam of peak energy 200 MeV and FWHM energy spread of 50 MEV is used to probe a region of plasma current filamentation.  

Jitter radiation solutions and spectral effects have thus far been largely developed and explored in application to astrophysical scenarios in which propagating shock fronts, magnetic reconnection, or other particle acceleration within the jetted outflow of highly relativistic collisionless plasma produce rapidly-evolving transient signals observed as gamma-ray bursts, X-ray flares, or other transient high-energy phenomena.\cite{mdv06,reyn+10}  Analytical treatments and PIC simulations indicate that such sources may produce an extended region of magnetic field turbulence generated by a relativistic Weibel-like filamentation instability. This instability has been called by a variety of names in the astrophysical literature and is often (incorrectly) referred to as the Weibel instability itself, but the basic mechanism is that small fluctuations in the magnetic field in a plane transverse to counterstreaming particles cause the particle streams to coalesce into current filaments which reinforce the initial field variations.\cite{mdv00,ged+10,bret+10}  We will here refer to it simply as a filamentation instability, with the understanding that we presume the filamentation has formed in this way.  

This filamentation instability can operate in a variety of scenarios and has been extensively studied in PIC simulations.\cite{nish+03,fred+04,bret+10}.  Much early work assumed such filamentation would occur at or upstream from a propagating shock front; however, recent results suggest that this type of instability may be much more ubiquitous, occuring also in non-shock plasmas with an appropriate anisotropy in particle distributions, and in outflow from sites of magnetic reconnection, among others.\cite{norreys+09}    

While simulations clearly show the production of filamentation by such an instability, this exact mechanism of filamentation remains challenging to identify in an experimental setting.  Certainly filamentary structures are seen in a wide variety of plasma scenarios, where they can degrade the efficiency of inertial confinement fusion and various laser and beam applications.  Plasma density filamentation is frequently observed in laser-plasmas (see [citations] for instance) but such observations have yet to be reliably studied with the simultaneous in-situ magnetic field and current measurements that would be required to identify them as a particular structure within a coherent theory.  The instabilities leading to this filamentation could provide valuable insights for laboratory astrophysics and high-energy density plasma physics, provided the instabilities can be adequately controlled and characterized.  The jitter radiation approach presented in this article could help with such studies.  Given a proposed form for the magnetic field wavenumber spectral distribution within a turbulent regime, we can calculate the resulting radiation spectra emitted towards various viewing directions for comparison with observations, allowing basic features of the magnetic field within the turbulent region to be confirmed without risk of disrupting or altering the instability with \textit{in situ} probes.  

In section \ref{sec:theory} we develop the equations for calculating the (small-angle) jitter radiation spectrum for any given magnetic field configuration, and then specifically for the anisotropic field distributions as generated a filamentation instability.  In section \ref{sec:results} we use these equations to calculate the jitter radiation spectra produced by particles moving in such an instability and by particles in an electron beam used to probe the filamentary region at various angles to the filamentation and separately characterize different components of the magnetic field.

\section{\label{sec:theory}Angle-resolved jitter radiation theory}

The angle-averaged jitter radiation emissivity equations have previously been developed and solutions presented for particles moving in the turbulent field produced by the relativistic Weibel instability within the internal shocks of a GRB jet.\cite{mdv00,mdv06,reyn+10} Here we present our derivation of the angle-resolved jitter emissivity such as may be used in laboratory applications.  
 	
The energy emitted per frequency per unit area for a relativistic charged particle with velocity $\bm{\beta}=\mathbf{v}/c$ that is undergoing an acceleration $\bm{\dot{\beta}}=d\mathbf{v}/cdt$ is given by the Leonard-Weichert equation \citep{LL,brau}:
	 \begin{equation}
	 \label{eq:LW}
	 \frac{d^2W}{d\omega d\Omega} = \frac{e^2}{4\pi^2c}\left|\int^{\infty}_{-\infty}dt\exp^{i\omega(t-\mathbf{\hat{n}}\cdot\mathbf{r}/c)}\frac{\mathbf{\hat{n}}\times\left[\left(\mathbf{\hat{n}}-\bm{\beta}\right)\times\bm{\dot{\beta}}\right]}{\left(1-\mathbf{\hat{n}}\cdot\bm{\beta}\right)^2}\right|^2
	 \end{equation}
For a highly relativistic particle such as we are considering, the velocity $\bm{\beta}$ is nearly constant and the time-dependence of the integrand is contained primarily in the acceleration $\bm{\dot{\beta}}$.  Using $\omega^{\prime}$ = $\omega(1-\mathbf{\hat{n}}\cdot\bm{\beta})$, we define $\bm{\xi} = \int\bm{\dot{\beta}}\exp^{i\omega^{\prime}t}dt$ as the Fourier transform of the particle's acceleration.  The above equation becomes:
\begin{equation}\label{eq:LWFour}
\frac{d^2W}{d\omega d\Omega} = \frac{e^2}{4\pi^2c}\frac{\left|\mathbf{\hat{n}}\times\left[\left(\mathbf{\hat{n}}-\bm{\beta}\right)\times\bm{\xi}\right]\right|^2}{\left(1-\mathbf{\hat{n}}\cdot\bm{\beta}\right)^2}
\end{equation}
We can define the components of $\bm{\beta}$ and $\bm{\xi}$ in the direction of an observer as $\beta_n = \mathbf{\hat{n}\cdot\bm{\beta}}$ and $\xi_n = \mathbf{\hat{n}}\cdot\bm{\xi}$ and define $\beta_{\bot}$ and $\xi_{\bot}$ as the magnitudes of the vector projection of $\bm{\beta}$ and $\bm{\xi}$ on the plane perpendicular to $\mathbf{\hat{n}}$.  We then find the angular spectral fluence to be:
\begin{equation}\label{eq:angform}
\frac{d^2W}{d\omega d\Omega} = \frac{e^2}{4\pi^2c}\left[\frac{\xi_n^2+\left|\bm{\xi}\right|^2}{(1-\beta_n)^2}-\frac{2\xi_n(\xi_n-(\bm{\beta}\cdot\bm{\xi}))}{(1-\beta_n)^3}+\frac{\beta_\bot^2\xi_n^2}{(1-\beta_n)^4}\right]
\end{equation}
The accelerations undergone by the particle, generated by the Lorentz force, will lie in the plane perpendicular to the particle's velocity.  Thus, $\bm{\beta\cdot\xi}$ = 0, and the above reduces to:
\begin{equation}\label{eq:angformsimp}
\frac{d^2W}{d\omega d\Omega} = \frac{e^2}{4\pi^2c}\left[\frac{\left|\bm{\xi}\right|^2}{(1-\beta_n)^2}+\frac{\xi_n^2(\beta^2-1)}{(1-\beta_n)^4}\right]
\end{equation}
For the case $\beta\parallel\mathbf{\hat{n}}$, this reduces to 
\begin{equation}\label{eq:headon}
\frac{d^2W}{d\omega d\Omega} = \frac{e^2}{4\pi^2c}\frac{\left|\xi_{\bot}\right|^2}{(1-\beta)^2}
\end{equation}
as expected.

To proceed further we need to determine the appropriate form for $\bm{\xi}$ relating the accelerations of the particle to the underlying magnetic field.  At a particular instant in time, a particle's acceleration $\bm{\dot{\beta}}=(e\beta/m\gamma)\mathbf{\hat{\bm{\beta}}}\times\mathbf{B}$ due to the Lorentz force is determined by the particle's velocity and the local magnetic field vector $\mathbf{B}$.  Recalling our earlier conditions, the particle velocity is nearly constant in magnitude and direction in the jitter regime.  Thus, the temporal evolution of the particle's acceleration reflects the propagation of the particle to a new position and thus becomes effectively an integral over a straight-line path of the particle through the turbulent magnetic field. 

\begin{equation}\label{eq:xi}
\bm{\xi}=\int\bm{\dot{\beta}}\exp^{i\omega^{\prime}t}dt=\left(\frac{e\beta}{m\gamma}\right)\hat{\bm{\beta}}\times\mathbf{B}_{\omega^{\prime}}
\end{equation}

where we have defined $\mathbf{B}_{\omega^{\prime}}=\int\mathbf{B}(\mathbf{r_o}+\bm{\beta} ct,t)\exp^{i\omega^{\prime}t}dt$.

\begin{eqnarray}\label{eq:xi-squared}
|\bm{\xi}|^2=\xi^*_i\xi^i&=&\left(\frac{e\beta}{m\gamma}\right)^2\left[|\mathbf{B}_{\omega^{\prime}}|^2-(\hat{\bm{\beta}}\cdot\mathbf{B}^*_{\omega^{\prime}})(\hat{\bm{\beta}}\cdot\mathbf{B}_{\omega^{\prime}})\right]\\
&=&\left(\frac{e\beta}{m\gamma}\right)^2(\delta_{\mu\nu}-\hat{\beta}_{\mu}\hat{\beta}_{\nu})W^{\mu\nu}
\end{eqnarray}
\begin{eqnarray}\label{eq:xi-n-squared}
(\bm{\xi}\cdot\hat{\mathbf{n}})^2 &=& \left(\frac{e\beta}{m\gamma}\right)^2\left[{\mathbf{B}^*_{\omega^{\prime}}}\cdot(\hat{\mathbf{n}}\times\hat{\bm{\beta}})\right]\left[{\mathbf{B}_{\omega^{\prime}}}\cdot(\hat{\mathbf{n}}\times\hat{\bm{\beta}})\right] \nonumber \\
&=&\left(\frac{e\beta}{m\gamma}\right)^2\left[\delta_{\mu\nu}(1-(\hat{\bm{\beta}}\cdot\hat{\mathbf{n}})^2)-\hat{\beta}_{\mu}\hat{\beta}_{\nu}-\hat{n}_{\mu}\hat{n}_{\nu}-(\hat{\bm{\beta}}\cdot\hat{\mathbf{n}})(\hat{\bm{\beta}}_{\mu}\hat{n}_{\nu}+\hat{\beta}_{\nu}\hat{n}_{\mu})\right]W^{\mu\nu}
\end{eqnarray}
where we have defined a tensor $W$ whose terms are products of the Fourier transformed magnetic field components and their complex conjugates, $W^{\mu\nu} = \mathbf{B}_{\omega^{\prime} }^{*\mu}\mathbf{B}_{\omega^{\prime}}^\nu$.  

We assume that the field and particle distributions are sufficiently homogeneous that the acceleration spectrum can be calculated from the statistically-averaged field in the radiating region of volume V.  That is, for a particular particle path across the radiating region, we presume that the exact starting position is irrelevant and exchange the integral of the field along a particular path for the volume-averaged magnetic field.  For simplicity, we also assume that the time $\Delta t_{prop}$ over which the particle is traversing the turbulent field region is much less than the dynamic time scale $\Delta T$ of the magnetic field turbulence, allowing us to treat the magnetic field distribution as static.  Thus, the magnetic field spectrum correlation tensor $W^{\mu\nu}$ becomes 
\begin{equation}\label{eq:Wavgform}
W^{\mu\nu} = \left<\mathbf{B}_{\omega^{\prime}}^{*\mu}\mathbf{B}_{\omega^{\prime}}^\nu\right> = (2\pi)^{-3}V^{-1}\int_{-\infty}^{\infty}\mathbf{B}_{\omega^{\prime}}^{*\mu} \mathbf{B}_{\omega^{\prime}}^{\nu}\delta(\omega^{\prime}-c\mathbf{k}\cdot\bm{\beta})d^3\mathbf{k}
\end{equation}
In equations \ref{eq:angformsimp}, \ref{eq:xi-squared}, \ref{eq:xi-n-squared}, and \ref{eq:Wavgform} we have not yet incorporated any particular geometry of the magnetic field, so these equations determine the jitter radiation spectra for any possible magnetic field distribution which varies on short enough scales to be within the jitter regime.

We now consider the geometry expected in a Weibel-type filamentation instability scenario, in which the magnetic field distribution along the filamentation axis (which we shall define here as $\mathbf{\hat{s}}$) is independent of the magnetic field distribution produced in the plane perpendicular to this axis.  Rewriting the magnetic field correlation tensor into a form which implicity contains this geometry, we obtain
\begin{equation}\label{eq:magfieldtensor}
\mathbf{B}_{\omega^{\prime}}^{*\mu} \mathbf{B}_{\omega^{\prime}}^{\nu}=(\delta_{\mu\nu}-\hat{s}_{\mu}\hat{s}_{\nu})\left|\mathbf{B}_{\mathbf{k}}\right|^2=(\delta_{\mu\nu}-\hat{s}_{\mu}\hat{s}_{\nu})f_{\perp}(k_\perp)f_{\parallel}(k_{\parallel})
\end{equation}
where $f_{xy}(k_{\perp})$ and $f_{z}(k_{\parallel})$ are the magnetic field wavenumber distributions in the directions transverse to and along the filamentation axis $\mathbf{\hat{s}}$ respectively.  Plugging this into equations \ref{eq:angformsimp}, \ref{eq:xi-squared}, and \ref{eq:xi-n-squared}, we obtain:
\begin{equation}\label{eq:angformfull}
\frac{dW}{d\omega d\Omega}=\frac{e^2}{4\pi^2c}G(\mathbf{\hat{s},\hat{n}},\bm{\beta})\left[\left(\frac{e\beta}{\gamma m}\right)^2\frac{1}{8\pi^3 V}\int|B_{\mathbf{k}}|^2\delta(\omega(1-\bm{\beta}\cdot\mathbf{\hat{n}})-c\mathbf{k}\cdot\bm{\beta})d^3\mathbf{k}\right]
\end{equation}
where we have defined the geometry and velocity dependent amplitude factor $G(\mathbf{\hat{s},\hat{n}},\bm{\beta})$ as 
\begin{equation}\label{eq:geofact}
G(\mathbf{\hat{s},\hat{n}},\bm{\beta}) = \left[\frac{1+(\mathbf{\hat{s}\cdot\hat{\bm{\beta}}})^2}{(1-\bm{\beta}\cdot\hat{\mathbf{n}})^2}-\frac{(\mathbf{\hat{s}\cdot\hat{\bm{\beta}}})^2+(\mathbf{\hat{s}\cdot\hat{n}})^2-2(\hat{\bm{\beta}}\cdot\mathbf{\hat{n}})(\mathbf{\hat{s}\cdot\hat{\bm{\beta}}})(\mathbf{\hat{s}}\cdot\mathbf{\hat{n}})}{(1-\bm{\beta}\cdot\mathbf{\hat{n}})^4}\right]
\end{equation}

\begin{figure}
\includegraphics{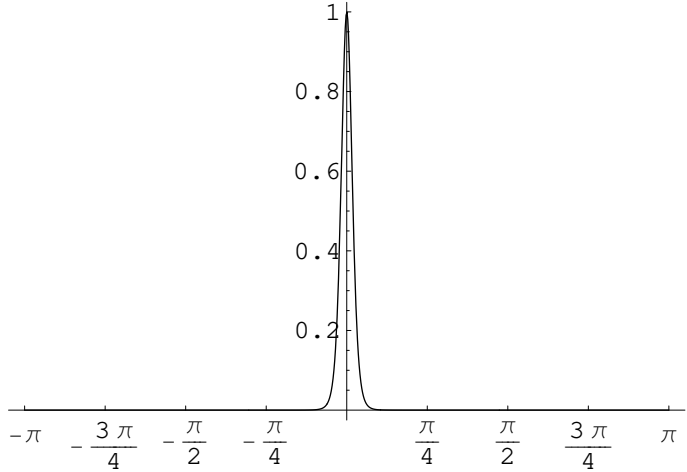}
\caption{\label{fig:geofact}The geometry dependent amplitude factor $G_{plane}(\beta,\theta,\alpha)$, where $\beta$ is the speed of a particle, $\theta$ is the angle between the particle's velocity and the filamentation axis $s$, and $\alpha$ is the angle between the particle's velocity and the direction of observation.}
\end{figure}

We can simplify our geometry slightly by considering the case in which $\mathbf{\hat{s},\hat{n}}$, and $\bm{\hat{\beta}}$ lie in the same plane.  In this case we can define the angles $\theta$ and $\alpha$ such that $\mathbf{\hat{s}}\cdot\hat{\bm{\beta}}=\cos\theta$, $\bm{\hat{\beta}}\cdot\mathbf{\hat{n}}=\cos\alpha$, and $\mathbf{\hat{s}\cdot\hat{n}}=\cos(\theta+\alpha)$.  In this planar case, Equation \ref{eq:geofact} becomes
\begin{eqnarray}\label{eq:geoplane}
G_{plane}(\beta,\theta,\alpha)& = & \frac{1}{(1-\beta\cos\alpha)^4}[\cos^2\alpha(1+\cos^2(\theta))+\cos^2\theta \nonumber \\
&&+\cos^2(\theta+\alpha)-2\cos\theta\cos\alpha\cos(\theta+\alpha)]
\end{eqnarray}
Figure \ref{fig:geofact} shows this factor plotted versus $\alpha$ for $\theta = 10^{\circ}$ and $\beta = 0.99$, normalized to its value at $\alpha$=0.  The strong peak of the emitted radiation in the forward direction is evident.  Despite the functional asymmetry of terms dependent on $\cos(\theta+\alpha)$ in \ref{eq:geoplane}, we find that the calculated emission is symmetric to well below 0.01\% for positive and negative $\alpha$.  

To solve for the power emitted per solid angle as a function of frequency, we make the substitution $|\mathbf{B_k}|^2 = f_{\perp}(\mathbf{k}_{\perp})f_{\parallel}(k_{\parallel})$ from Equation \ref{eq:magfieldtensor} into Equation \ref{eq:angformfull}.  
\begin{equation}\label{eq:lwang}
\frac{d^2W}{d\omega d\Omega} = \frac{e^4\beta^2}{4\pi^2 \gamma^2 m^2 c}\frac{G(\mathbf{\hat{s}},\mathbf{\hat{n}},\bm{\beta})}{8\pi^3V}\int f_{\parallel}(k_\parallel)f_{\perp}(k_\perp)\delta\left(\omega^{\prime}-c\mathbf{k}\cdot\bm{\beta} \right)d^2k_{\perp}dk_{\parallel}
\end{equation}
In most situations, the magnetic field spectrum along and transverse to the filamentation axis cannot be known from first principles but may be approximated from some knowledge of the relevant parameters.  The magnetic field spectrum transverse to the filamentation axis has been shown to rise and then drop after a scale of order the plasma skin depth \citep{fred+04,ml99}.  The filamentation instability itself produces this transverse field distribution independently at each point along the filamentation axis.  As in \citet{mdv06} and \citet{reyn+10}, we define a geometry where the filamentation axis (i.e., the orientation of the currents from the original counterstreaming particle populations) is along the $z$-direction $(\mathbf{\hat{s}}\parallel\mathbf{z})$ and the cross-section across the filaments can be taken as lying locally in the $xy$-plane.  To calculate the resulting spectra, we have used a general spectral form that may be parameterized as   
\begin{equation}\label{eq:perpfspec}
f_{xy}(k_{\perp})=\frac{k_{\perp}^{2\alpha_{\perp}}}{\left( \kappa_{\perp}^2+k_{\perp}^2\right)^{\alpha_{\perp}+\beta_{\perp}}},
\end{equation}
where $\kappa_{\perp}$, $\alpha_{\perp} > 0$, and $\beta_{\perp} > 0$ are free parameters controlling the spectral break and the soft and hard spectral indices, respectively.  In this equation $k_{\perp} = (k_x^2 + k_y^2)^{1/2}$ is the cross-filamentation magnetic field wavenumber. Similarly, we characterize the magnetic field spectrum along z as
\begin{equation}\label{eq:pllfspec}
f_{z}(k_{\parallel})=\frac{k_{\parallel}^{2\alpha_{\parallel}}}{\left( \kappa_{\parallel}^2+k_{\parallel}^2\right)^{\alpha_{\parallel}+\beta_{\parallel}}},
\end{equation}
where $\kappa_{\parallel}$, $\alpha_{\parallel} > 0$ and $\beta_{\parallel} > 0$ are independent free parameters.  We have defined our field spectra (Equations \ref{eq:perpfspec} and \ref{eq:pllfspec}) so as to make variables $k$ and $\kappa$ dimensionless values, expressed in terms relative to a wavenumber unit $k_0$.  This will result in a normalizable factor $B_{p0} k_0^{-2\beta_p}$ contributed to Equation \ref{eq:lwang} by each spectral distribution, where $p$ simply designates that we are considering either the parallel or perpendicular component of wavenumber.  From the delta function in Equation \ref{eq:lwang} this means our frequencies will be defined in terms of units $\omega_0 = ck_0$.  The basic asymptotic behavior of the equations for the field spectra is summarized as
\begin{equation}\label{eq:fldasympt}
f(k_p)\propto\left\{
 \begin{array}
  {l@{\qquad}l}
	k_p^{2\alpha_p} ,& \mbox{if  } k_p<<\kappa_p,\\	
	k_p^{-2\beta_p} ,& \mbox{if  } k_p>>\kappa_p.
	\end{array}
	\right. 
\end{equation}

For the limiting cases in which the beam is aligned directly along ($\theta=0^{\circ}$) or transverse ($\theta=90^{\circ}$) to the filamentation axis $\bm{\hat{s}}$ the term $\bm{k\cdot\beta}$ within the delta function in Equation \ref{eq:lwang} reduces to a function of a single component of $k$ (with appropriate choice of the $x$-axis) so that the integral over the delta function amounts to a simple substitution for $k_x$ or $k_z$ as a function of $\omega$, $\theta$, and $\beta$.  This results in a spectrum whose frequency dependence is determined only by the distribution of the magnetic field along (for $\theta = 0^{\circ}$) or perpendicular (for $\theta=90^{\circ}$) to the filamentation axis.  For a more general orientation of the beam at some angle $0^{\circ}<\theta<90^{\circ}$ relative to the filamentation axis, the substitution for $k_x$ or $k_z$ includes the other wavenumber component and the frequency dependence of the resulting spectrum is influenced by both magnetic field distributions.  

\section{\label{sec:results}Results and Implications}

A typical laser-plasma experiment for studies of the filamentation (Weibel-type) instabilities generally involves a strong beam that produces anisotropy in the plasma distribution function (PDF) which then drives the instability. The parallel (beam) direction is the direction of the PDF anisotropy, and it is the direction in which the current filaments are oriented. Radiation is produced by the plasma particles propagating in the magnetic fields of the filaments and it can be used to determine their properties. Alternatively, one can also launch a probe beam of electrons into the plasma and use the radiation produced by these particles to diagnose the magnetic fields. The latter scenario is more demanding in terms of engineering difficulties, but it should provide a more powerful diagnostics. Indeed, one generally have a better control on the energy distribution of the radiation-producing (probe beam) particles as well as one can probe a variety of incident angles $\theta$ relative to the filamentation direction. In both cases we base our radiating particle population on a quasi-monoenergetic electron beam consisting of a Gaussian distribution of particle energies, with peak energy of 200 MeV and a full-width at half maximum of 50 MeV.  

For simplicity and in the absence of any particular plasma parameters governing our magnetic field distribution, we have chosen to use the same spectral parameters for the magnetic field distributions along and transverse to the filamentation axis.  As we have already mentioned, the spectral shape is determined by the components of the magnetic field distribution along the beam so if there are distinctly different parameters for the magnetic field distribution along and perpendicular to the filamentation axis, the influence of these field parameter anisotropies can easily be isolated by looking at the emission from electrons moving along or transverse to the filamentation.  We have used arbitrary-yet-reasonable selected values for the field spectral peak $\kappa_{\perp} = \kappa_{\parallel} = 10$, low-wavenumber spectral indices of $\alpha_{\perp} = \alpha_{\parallel} = 2$, and high-wavenumber spectral indices of $\beta_{\perp} = \beta_{\parallel} = 1.5$.  Our $\kappa_p$ correspond to the peak of each field distribution in wavenumber, defined to be in terms of units $k_0 = \omega_0/c$.  Defining our units in terms of the field correlation length $\lambda_{B\perp}$ in the direction transverse to the filamentation axis (the transverse field distribution being better developed by theory), we can define $k_0 = 10/\lambda_{B\perp}$ so that  $\kappa_{\perp} = \kappa_{\parallel} = 1/\lambda_{B\perp}$.  We then have frequency units $\omega_0 = 10c/\lambda_{B\perp}$.    

We find that in either case the emissivity is strongly peaked in the forward beam direction as expected.  The spectral distribution of the generated spectrum is dependent primarily on the distribution of the magnetic field wavenumber's component along the beam.  Thus, the spectrum produced by the instability-generating beam contains details about the magnetic field distribution along the filamentation axis, as in subsection \ref{ss:scenario1}.  Obtaining information about the transverse magnetic field distribution requires a secondary beam to probe the instability region at angles transverse to the current filamentation, as we demonstrate in subsection \ref{ss:scenario2}.  

\subsection{\label{ss:scenario1}Radiation from instability-generating beam}

In this scenario, we consider the radiation produced by electrons in the same beam that produces the filamentation instability, slightly after the formation of the instability itself.  The development of the filamentation instability is such that the filamentation axis $\mathbf{\hat{s}}$ will be aligned with the direction of the generating beam $\mathbf{\hat{r}}_{beam} \parallel \bm{\hat{\beta}}$.  In this case, we find $\mathbf{k}\cdot\bm{\hat{\beta}} = k_{\parallel}$ and consequently, the components of $\mathbf{k}$ perpendicular to the filamentation axis are eliminated from the delta function in Equation \ref{eq:lwang}.  The frequency dependence of the resulting radiation spectrum is thus obtained solely from the magnetic field distribution along the filamentation axis, $f_{\parallel}(k_{\parallel})$. 

Beam divergence will contribute a small influence from the transverse magnetic field to the spectrum emitted in the beam forward direction by the full particle distribution, but such contributions are strongly limited by the anisotropy of the particle distribution and the relativistic beaming factor as was shown in Figure \ref{fig:geofact}. For simplicity, we also neglect here the self-consistent modifications of the beam as it propagates through the instability.  

\begin{figure}
\includegraphics[width = 0.90\textwidth]{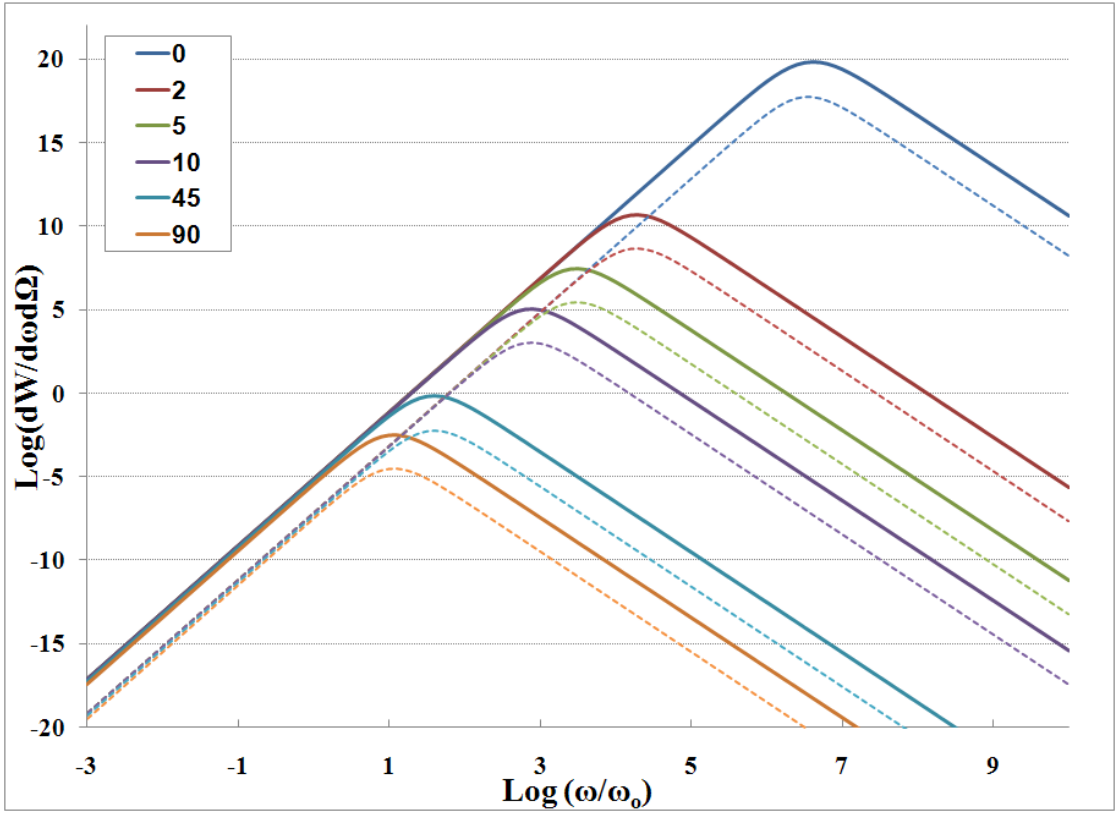}
\caption{\label{fig:theta0}The spectrum produced by electrons moving parallel to the filamentation axis, when viewed from different angles of observation $\alpha$, defined as the angle between the direction to the observer and the direction in which the beam propagates along the current filaments.  Solid lines show the emission generated by the full distribution of beam particle energies; dashed lines show emission produced by an individual particle with the beam's peak energy.  The maximum emission and highest peak frequency are produced in the beam's forward direction ($\alpha=0$).  As it progresses to larger viewing angles (shown are $\alpha=2^{\circ},5^{\circ},10^{\circ},45^{\circ}$, and $90^{\circ}$), the spectrum maintains its overall shape but dims overall and softens in peak energy.}
\end{figure}

Figure \ref{fig:theta0} shows the resulting radiation spectrum produced by electrons within the same beam that produced the instability, when the spectrum are obtained from various angles $\alpha$ = $0^{\circ}, 2^{\circ}, 5^{\circ}, 10^{\circ}, 45^{\circ}$, and $90^{\circ}$ relative to the identical direction of the beam and the filamentation axis ($\mathbf{\hat{r}}_{beam} \parallel \bm{\beta} \parallel \mathbf{s}$, or equivalently $\theta=0^{\circ}$).  The dashed line shows the spectrum produced by a particle of the electron beam's peak energy and the solid line shows the approximate overall spectrum, calculated as a weighted sum over the spectra produced by particles of different $\gamma$.  The values on the vertical axis are essentially arbitrary, as they depend on the average magnetic field strength generated by the instability, the electron beam particle density, and the total volume of the radiating region in which the beam intersects with the instability.  Our horizontal axis is normalized to $\omega_0 = ck_0$, which we have defined above as $10c\lambda_{B\perp}$.  
 
\subsection{\label{ss:scenario2}Radiation from probe beam of varying incident angles}

\begin{figure}
\includegraphics[width=0.90\textwidth]{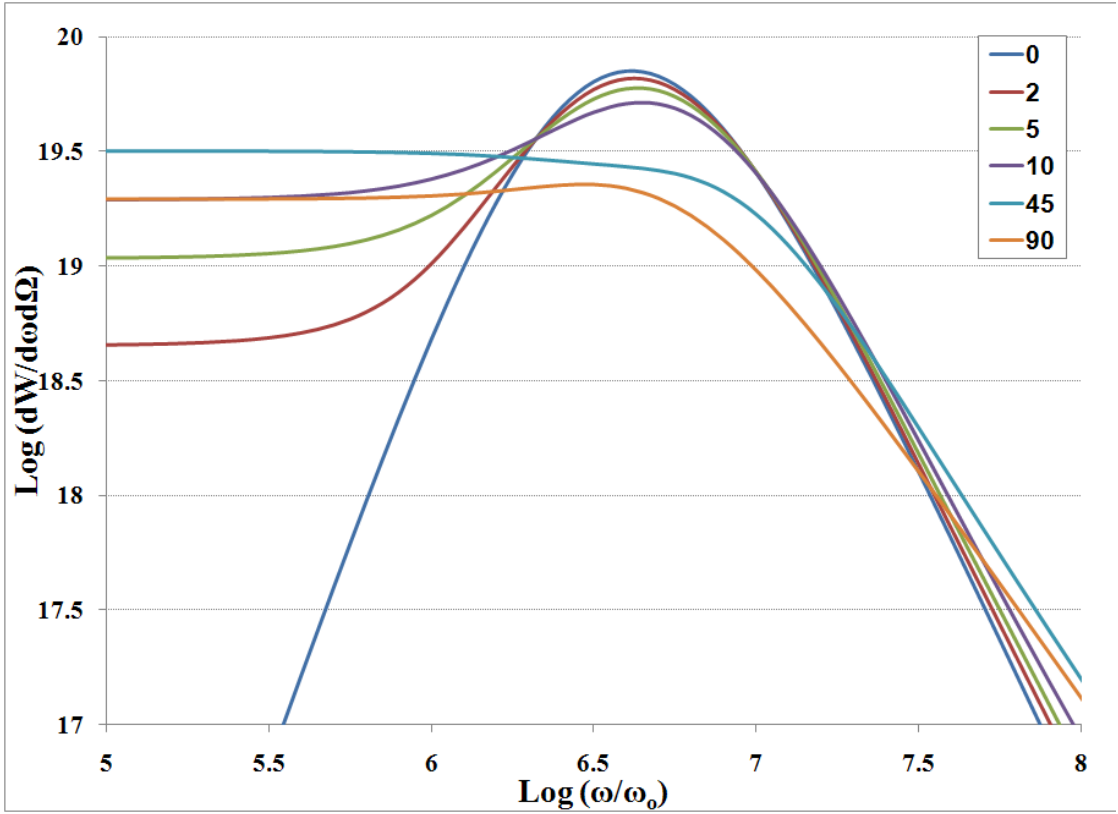}
\caption{\label{fig:thetacomp}Calculated jitter radiation spectra such as produced a representative quasimonoenergetic beam of particles probing the filamentation instability region at angles $\theta$ = $0^{\circ}, 2^{\circ}, 5^{\circ}, 10^{\circ}, 45^{\circ}$, and $90^{\circ}$ relative to the filamentation axis.  The spectrum notably changes from a peaked form with low-energy spectral index of approximately 2 at $\theta = 0$, through peaked forms with a second low-energy break at $\theta = 2^{\circ},5^{\circ}$, and $10^{\circ}$, to forms with a break (but no distinguishable peak) and a flat low-energy spectral index of 0 at $\theta = 45^{\circ}$ and $90^{\circ}$.}
\end{figure}

Exploring the magnetic field distribution transverse to the filamentation axis requires obtaining and analyzing the radiation produced by particle populations moving in these directions.  We consider that the electron beam used to generate the instability may be split prior to entering the plasma (or a second beam generated) and redirected to probe the instability from other angles $\theta$ between the filamentation axis and the radiating beam.  Since the peak emissivity will still be in the radiating particle populations forward direction, the radiation spectra would ideally be obtained with movable detectors that can be positioned in or close to the direction of the beam's path through the filamentation region (outside of the filamentary region the beam may be deflected so as not to hit the detectors).  Figure \ref{fig:thetacomp} shows the spectrum for incident beam angles of $\theta$ = $0^{\circ}, 2^{\circ}, 5^{\circ}, 10^{\circ}, 45^{\circ}$, and $90^{\circ}$ when viewed from the beam's forward direction ($\alpha=0$).  As in the previous subsection, the vertical normalization is essentially arbitrary, but the horizontal normalization is defined in terms of $\omega_0 = ck_0 = 10c\lambda_{B\perp}$, where $\lambda_{B\perp}$ is the correlation length of the magnetic field distribution transverse to the filamentation. The case $\theta=0^{\circ}$ is of course equivalent to that considered in subsection $\ref{ss:scenario1}$ if there is no variation in the energy distribution between the probe beam and the beam that generated the instability.

\begin{figure*}
\centering
\subfigure[]{\includegraphics[width = 0.48\textwidth]{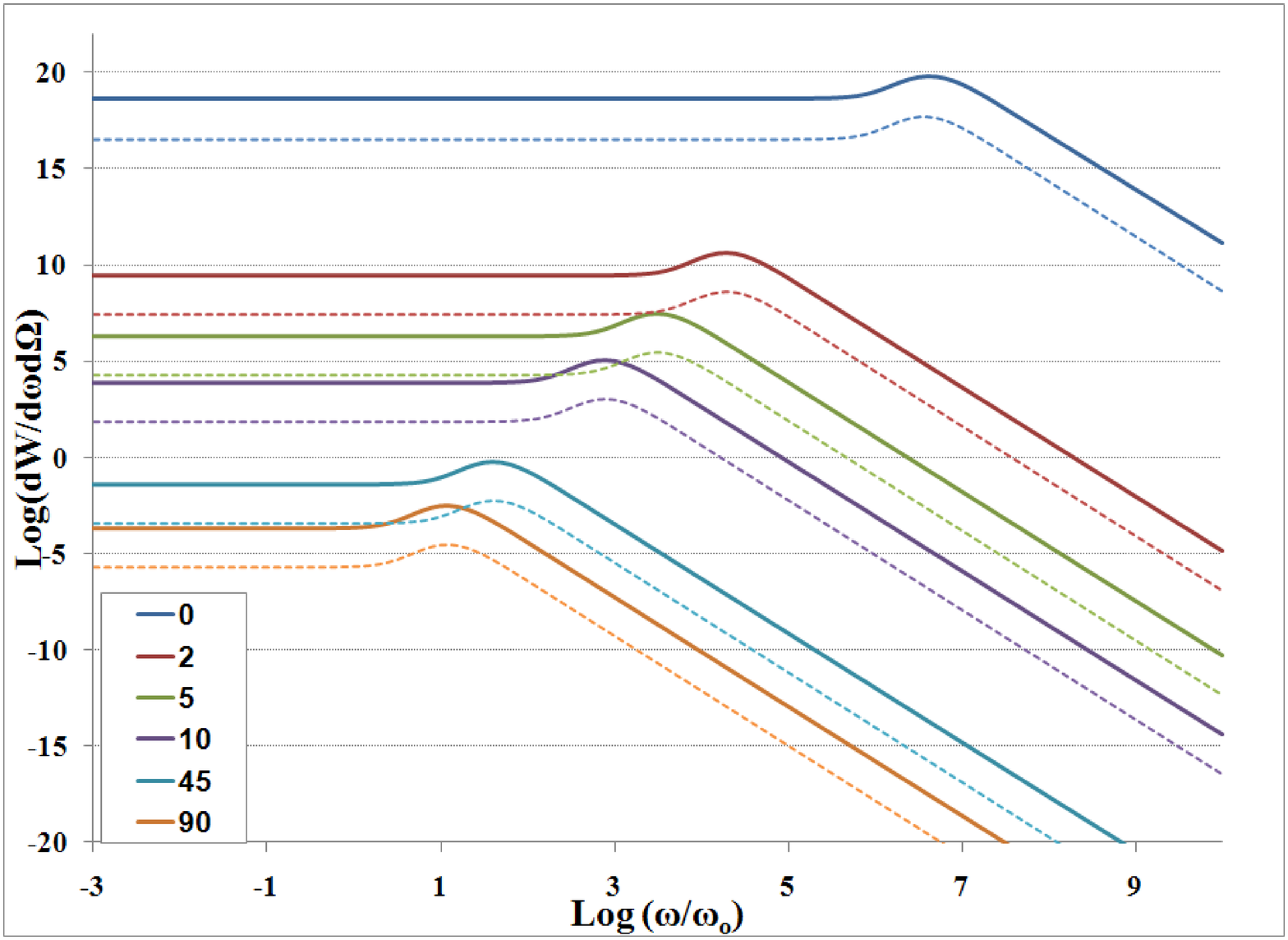}} 
\subfigure[]{\includegraphics[width = 0.48\textwidth]{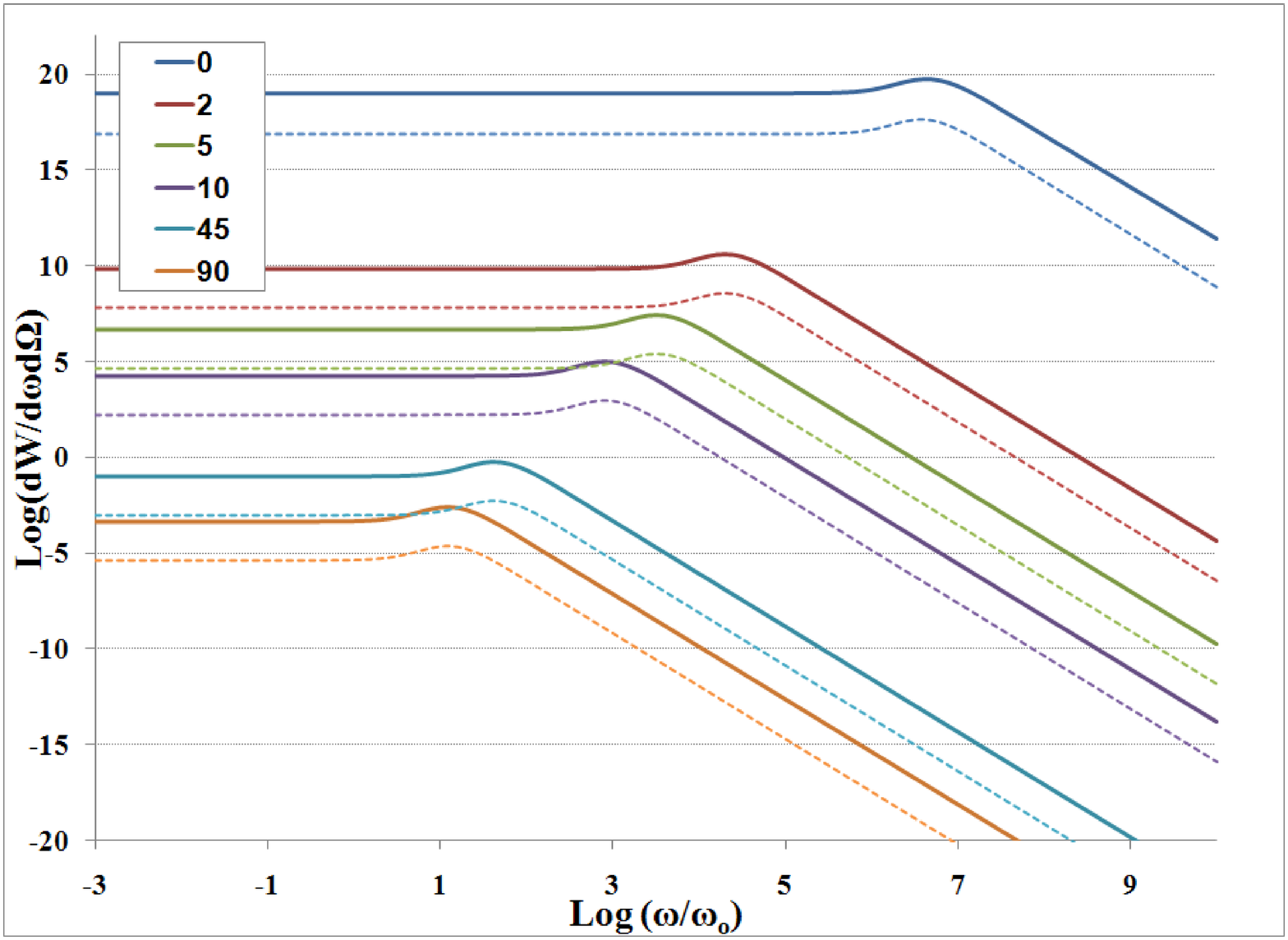}}\\
\subfigure[]{\includegraphics[width = 0.48\textwidth]{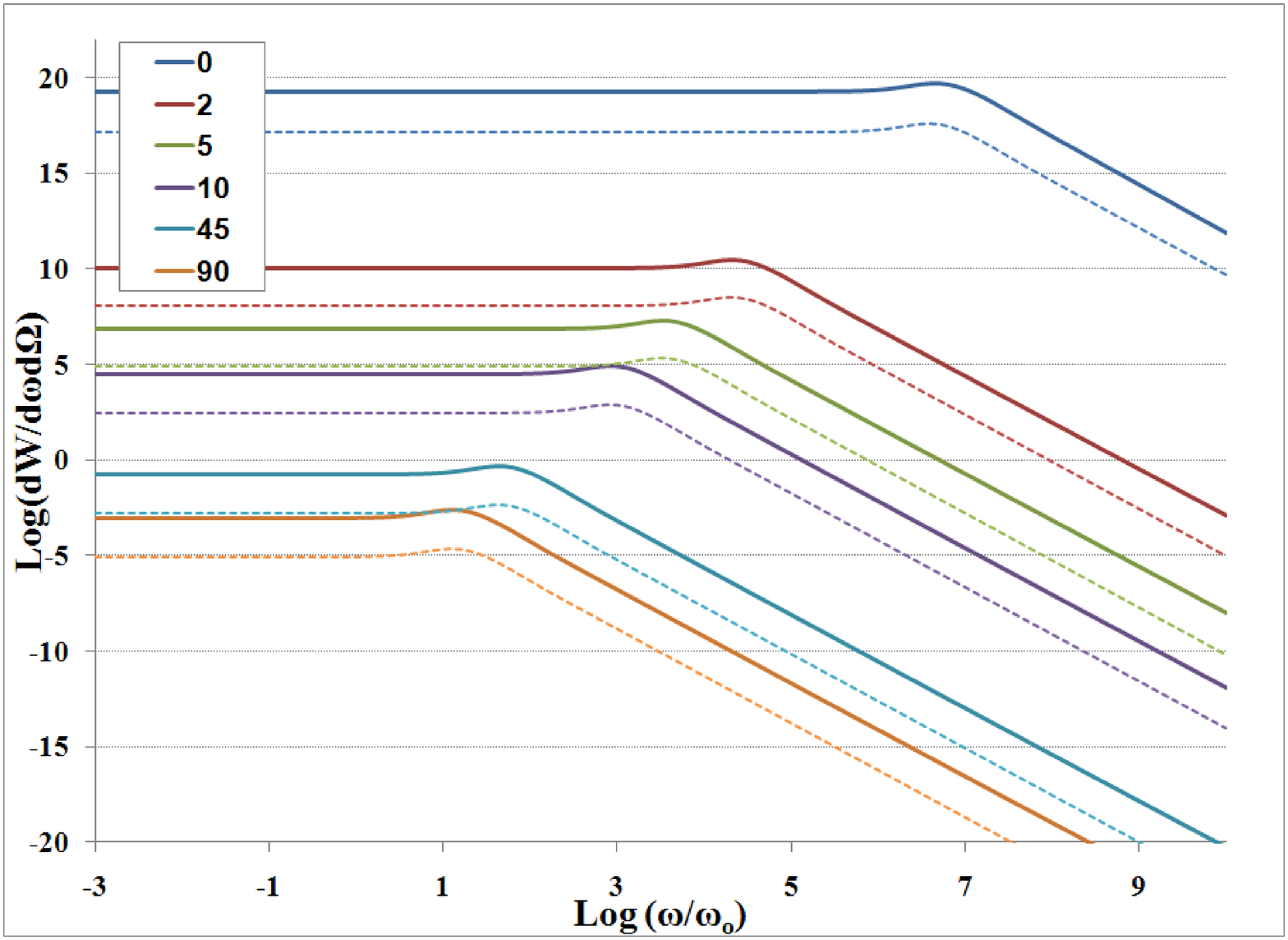}}
\subfigure[]{\includegraphics[width = 0.48\textwidth]{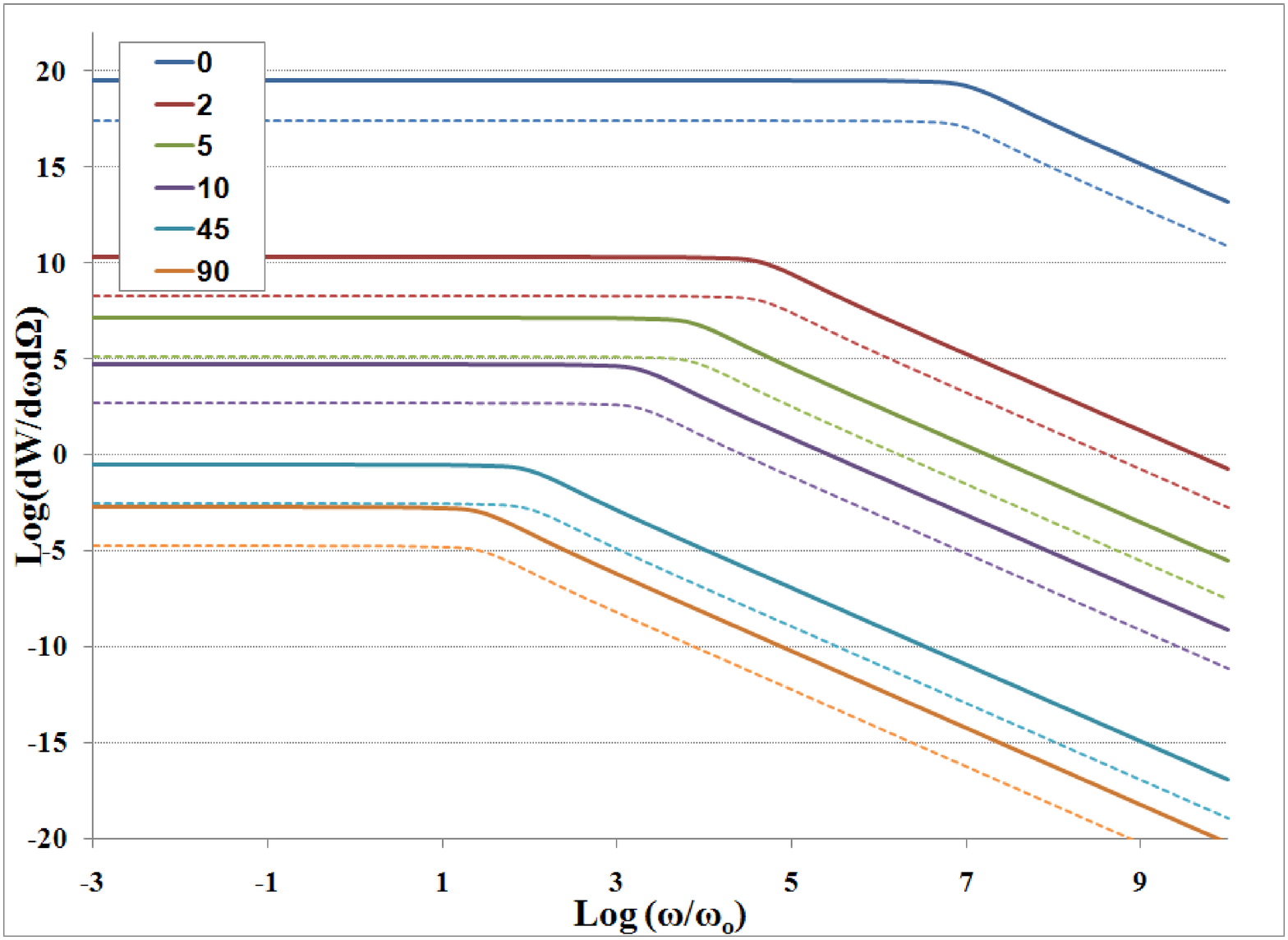}}\\
\caption{\label{fig:thetavar}Variation in the jitter radiation spectrum produced by a quasimonoenergetic beam (solid lines) and an individual particle of the beam's peak energy (dotted lines).  Each figure shows results for a particular angle $\theta$ between the probe beam and the instability's filamentation axis, when viewed from different viewing angles $\alpha = 2^{\circ}$ (a), $5^{\circ}$ (b), $10^{\circ}$ (c), and $45^{\circ}$ (d), all measured from the beam direction $\bm{r}_{beam}$ in the same plane as $\theta$.}
\end{figure*}

For each $\theta$, we have also calculated the spectrum for several different viewing angles $\alpha = 2^{\circ}, 5^{\circ}, 10^{\circ}, 10^{\circ}, 45^{\circ}$ and $90^{\circ}$, where $\alpha$ is again defined as the angle between the direction to an observer and the unit vector $\mathbf{\hat{r}}_{beam}$ along the probe beam, as measured in the same plane as the angle $\theta$ on the far side of $\mathbf{r}_{beam}$ from $\mathbf{\hat{s}}$.  Figure \ref{fig:thetavar} shows the results for probe beam incident angles of $\theta = 2^{\circ}, 5^{\circ}, 10^{\circ}, 45^{\circ}$ respectively.  We have omitted the $\theta=90^{\circ}$ case from Figure \ref{fig:thetavar} has been omitted since at these scales it appears nearly identical to the $\theta=45^{\circ}$ case (see Figure \ref{fig:thetacomp}). As before, solid lines show the weighted sum over our distribution of $\gamma$, while dashed lines show the spectrum produced by an individual particle at the peak value of $\gamma$ (here a representative 200/0.511 Mev).  The spectrum from an individual particle and the full distribution differ notably in amplitude but not significantly in overall shape.  For all of the probe beam angles, the spectrum is strongest when viewed along the probe beam (viewing angle $\alpha = 0^{\circ}$), with amplitudes decreasing rapidly by several orders of magnitude even when viewed at comparatively small viewing angles of $\alpha = 2^{\circ}, 5^{\circ},$ and $10^{\circ}$ (in keeping with the relativistic beaming we expect).  While the overall spectral shape is unchanged, the viewing angles $\alpha$ also result in a shift of the spectrum to shorter wavelengths $\omega$.  This shift can be determined analytically from equation \ref{eq:angformfull}, from which we find that for the case $\theta = 0^{\circ}$ a spectral feature located at $\log{\omega_p}$ in the $\alpha=0^{\circ}$ will be shifted to  $\log\omega_p-\log((1-\beta\cos\alpha)/(1-\beta))$ when viewed at other angles $\alpha$. 

\section{Conclusions}

We have developed the jitter radiation equations appropriate for analyzing the radiation produced within laboratory laser-plasma interactions in which relativistic particles radiate as they move through magnetic field turbulence without being substantially deviated from a path along a particular line of sight. Because the resulting jitter radiation spectrum depends directly upon the magnetic field distribution along the particle's path, it has valuable potential as a non-invasive diagnostic for turbulent laboratory plasmas.  Although we present results for a particular application in this paper, the approach developed here can be applied much more generally.  Given a three-dimensional spectral distribution of the magnetic field wavenumber, we can directly calculate the resulting radiation spectrum in the jitter regime from particle beams propagating through this field distribution.  

In this paper we have demonstrated this application of jitter radiation to the Weibel-like filamentation instability, which is of great interest for both astrophysical and laboratory plasmas.  We calculate the radiation spectrum as would be produced by electrons in a quasi-monoenergetic beam that generates the filamentation instability as it impinges on a plasma.  We demonstrate that the resulting radiation spectrum depends directly upon the magnetic field distribution along the filamentation axis, with very little sensitivity to the distribution of magnetic wavenumber transverse to this axis.  The resulting spectrum is seen to be distinctly harder than the synchrotron case, with a low-frequency spectral index of 2.  The spectrum peaks at higher frequencies in the beam's forward direction, and then diminishes in overall intensity and in peak frequency when viewed from other angles.  

We also consider a scenario in which a second quasi-monoenergetic electron beam probes the existing instability at varying incident angles relative to the instability's filamentation axis.  Taking advantage of this method's control of density and orientation of the radiating particles, we can vary the angle to isolate the spectral influence of either of the independent magnetic field distributions along and transverse to the filamentation axis.  As we vary the probe beams incident angle, we progress from a strongly peaked spectral form to peaked forms that have a low-frequency flattening, to unpeaked ``broken power-law type'' forms with low-frequency spectral index of 0 and a single break to the high-frequency spectral index.  

This work has been supported by NSF grant AST-0708213, NASA ATFP grant NNX-08AL39G and DOE grant DE-FG02-07ER5494.  We also gratefully acknowledge the computer system expertise of our colleague Sriharsha Pothapragada.

\end{document}